\documentclass[11pt,a4paper]{article}

\usepackage[utf8]{inputenc}
\usepackage[margin=1.4in]{geometry}
\usepackage{graphicx,wrapfig}
\usepackage[draft]{todonotes}
\usepackage{amsmath,amssymb}
\usepackage{physics,braket}
\usepackage{mathrsfs,breqn}
\usepackage{hyperref}

\numberwithin{equation}{section}
\numberwithin{figure}{section}
\title{ Integrable deformations of the flat space sigma model}

\author{\vspace{-0.25cm}\\
   Khalil Idiab
  \vspace{0.25cm}\\
 \footnotesize \em Institut f\"ur Mathematik und Institut f\"ur Physik, Humboldt-Universit\"at zu Berlin, \\
   \footnotesize \em IRIS Geb\"aude, Zum Grossen Windkanal 6, 12489 Berlin, Germany
 }
 \date{\small\url{khalil.idiab@physik.hu-berlin.de}}

\begin{document}
 \maketitle
   \pagenumbering{gobble}
   \thispagestyle{headings}
  \markright{\hfill \normalfont\texttt{HU-EP-24/23} }
\begin{abstract}

  We explore a deformation of the flat space symmetric space sigma model action. The deformed action is designed to allow a Lax connection for the equations of motion, similar to the undeformed model. For this to work, we identify a set of constraints that the deformation operator, which is incorporated into the action, must fulfil. After defining the deformation, we explore simple solutions to these constraints and describe the resulting deformed backgrounds. Specifically, we find flat space in Cartesian coordinates with arbitrary constant $H$-flux or linear $H$-flux in a light cone coordinate. Additionally, we find the Nappi-Witten background along with various Nappi-Witten-like backgrounds with near arbitrary constant $H$-flux. Finally, we discuss the symmetries of the deformed models, finding that the deformed symmetries will always include a set of symmetries that in the undeformed limit becomes the total set of translations.

\end{abstract}
\tableofcontents
\newpage

\pagenumbering{arabic}
\setcounter{page}{2}
\section{Introduction}
Integrable sigma models have played an important role in the exploration of holography and the AdS/CFT correspondence. A major focus in this area has been on $\mathrm{AdS}_{n}$ and $\mathrm{S}^{n}$, which are studied as symmetric space sigma models. These algebra-based models have been crucial in understanding the integrability structure of free strings on curved backgrounds and have provided insights into non-perturbative aspects of AdS/CFT \cite{Metsaev:1998it, Arutyunov:2009ga, Bena:2003wd, Beisert:2010jr, Bombardelli:2016rwb}.

In this realm, Yang-Baxter deformations of integrable sigma models and particularly of the symmetric space sigma model have been extensively studied \cite{Klimcik:2002zj,Klimcik:2008eq, Delduc:2013fga,Delduc:2013qra,Kawaguchi:2014qwa,vanTongeren:2015soa,Hoare:2021dix}. These are a class of deformations that preserve the integrability structure of the sigma models. Some of these deformations have been linked to various geometric interpretations such as TsT-transformations \cite{Osten:2016dvf}, non-abelian T-duality \cite{Hoare:2016wsk,Borsato:2018idb} and twisted boundary conditions \cite{Frolov:2005dj,Vicedo:2015pna,vanTongeren:2018vpb,Borsato:2021fuy}. Algebraically, some of these deformations are linked to Drinfel'd twists \cite{Vicedo:2015pna, vanTongeren:2015uha, vanTongeren:2016eeb, vanTongeren:2021jhh}. It is conjectured that their dual gauge theory side corresponds to non-commutative field theory \cite{vanTongeren:2015uha, vanTongeren:2016eeb}, and some of these theories have recently been constructed in \cite{Meier:2023kzt,Meier:2023lku}. Due to applications within holography, the study of Yang-Baxter deformations has primarily been focused on the maximally symmetric $\mathrm{AdS}_{5}$ and $\mathrm{S}^{5}$. To gain further insights into the nature of Yang-Baxter deformation, it is valuable to study them for different models or target spaces. Flat space provides an advantageous arena for such studies, as it contains simple deformations that may even be exactly solved \cite{Idiab:2024bwr}.

In addition to expanding our understanding of Yang-Baxter deformations, it is an interesting pursuit to find other integrable deformations. In this paper, we present an example that was orginally inspired by studying the Lax connection of the Nappi-Witten model represented as a Yang-Baxter deformation of flat space. Specifically, we consider a deformation of the symmetric space sigma model of flat space that retains its Lax integrability. This is achieved by inserting a linear operator into the action, which crucially maintains the appropriate coset gauge invariance. We then identify the conditions this operator must fulfill for our Lax ansatz to be compatible with the equations of motion. Subsequently, we consider simple solutions to these constraints and determine their associated deformed backgrounds. The full space of deformations remains to be explored. Furthermore, the possibility of finding similar integrable deformations of other (non-flat) symmetric spaces remains unexplored but intriguing.

The structure of this paper is organized as follows. First, we recapitulate the symmetric space sigma model specifically from the perspective of flat space. After stating the Lax connection for this model, we discuss possible modifications to this Lax connection to accommodate potential deformations. In the second section, we provide an example of an action that fits the deformation scheme proposed in the previous section. Next, we consider simple examples of deformed backgrounds that fulfill the necessary constraints for Lax integrability. In the final section, we study the symmetries of the deformed backgrounds.

\section{Flat space symmetric space sigma model}
The worldsheet action of a string in flat space can be represented in various ways. For our purposes, the symmetric space sigma model is the relevant formulation. In this section, we will review this formulation to set our conventions. For more details, see \cite{Idiab:2022dil}.
We consider flat space as the symmetric space of the Poincar\'e algebra
\begin{align}\label{eq:lorentzcommutators}
  [m_{\mu\nu},m_{\rho\sigma}]&=\eta_{\mu\sigma} m_{\nu\rho}+\eta_{\nu\rho}m_{\mu\sigma}-\eta_{\mu\rho}m_{\nu\sigma}-\eta_{\nu\sigma}m_{\mu\rho},\\
  [m_{\mu\nu},p_{\rho}]&=\eta_{\nu\rho} p_{\mu}-\eta_{\mu\rho}p_{\nu}, \qquad \qquad [p_{\mu},p_{\nu}]=0,
\end{align}
where $m_{\mu\nu}$ are Lorentz transformations, $p_{\mu}$ are translations and $\eta_{\mu\nu}$ is the Minkowski metric. This algebra is crucially $\mathbb{Z}_2$ graded
\begin{align}
\mathfrak{g}=\mathfrak{g}^{(0)} \oplus \mathfrak{g}^{(1)},&& [\mathfrak{g}^{(i)},\mathfrak{g}^{(j)}]\subset \mathfrak{g}^{(i+j\hspace{-0.15cm}\mod 2)},
\end{align}
with translations in grade $1$ and Lorentz transformations in grade $0$. To write an action for this model, we need a Lorentz invariant non-degenerate bilinear form, which we take to be
\begin{align}\label{eq:bilinearform}
  \ip{m_{\mu\nu}}{m_{\rho\sigma}}  = \eta_{\mu\sigma}\eta_{\nu\rho}-\eta_{\mu\rho}\eta_{\nu\sigma},\\
  \ip{m_{\mu\nu}}{p_{\rho}} =0, \quad \ip{p_{\mu}}{p_{\nu}}=\eta_{\mu\nu}.
\end{align}
These properties of the Poincar\'e algebra make it suitable for the symmetric space sigma model formulation. To construct the action, we consider a coset field $g$ living on the two dimensional worldsheet $\Sigma$. The action is written in terms of the Maurer-Cartan one-form $A=-g^{-1}dg$,\footnote{The Maurer-Cartan form can be expressed in terms of worldsheet one-form coefficients $A=A_{\alpha}d\sigma^{\alpha}$ or target space one-form coefficients $A=A_{\mu}dx^{\mu}$, where $\sigma^{\alpha}$ are coordinates on the worldsheet and $x^{\mu}$ are target space coordinates.}
\begin{align}
S[g]= \frac12 \int_{\Sigma} \ip{A}{\star \mathcal{P} A},
\end{align}
where the inner product contains an implicit wedge product. The algebra projector $\mathcal{P}: \mathfrak{g}\rightarrow\mathfrak{g}^{(1)}$, projects to the translations. The advantage of this formulation is that it manifests all the symmetries and is coordinate-independent. For example, using the coset representative $g=e^{x^{\mu}p_{\mu}}$, the action becomes the flat space string in Cartesian coordinates, while other coset representatives will yield the same background in different coordinate systems. Concomitantly, the equations of motion can be expressed in a coordinate independent way
\begin{align}
\label{eq:2}
\mathcal{E}=d\star \mathcal{P} A- A \wedge \star \mathcal{P}A - \star \mathcal{P}A \wedge A,
\end{align}
or in terms of worldsheet one-form coefficients
\begin{align}
\mathcal{E}=\partial_{\alpha}\mathcal{P}A^{\alpha}-[A_{\alpha},\mathcal{P}A^{\alpha}].
\end{align}
 We can find a Lax connection for these equations for generic symmetric spaces with the ansatz
\begin{align}
  L(z)=A+ \ell_{1}(z) \star \mathcal{P} A+ \ell_{2}(z)  \mathcal{P} A.
\end{align}
In general, this ansatz is valid under the constraints $\ell_{1}\neq 0$ and $\ell_{1}^{2}+\ell_{2}^{2}+2\ell_{2}=0$. However, since grade $1$ elements commute for the Poincar\'e algebra, only the first constraint is necessary. This allows us to use the simple form
\begin{align}
  \label{eq:undeformedLax}
L(z)=A+ z \star \mathcal{P} A,
\end{align}
where $z$ is a spectral parameter. The curvature of this Lax connection is
\begin{align}
  \mathcal{F} L &= dL - L \wedge L,\\
  &=\mathcal{F}A+ z \left( d \star \mathcal{P} A -  A \wedge  \star \mathcal{P}A -\star \mathcal{P}A \wedge A \right)- z^{2} \star \mathcal{P}A \wedge \star \mathcal{P}A.
\end{align}
The Maurer-Cartan form is automatically flat $\mathcal{F}A=0$, and for the Poincar\'e algebra $\star \mathcal{P}A \wedge \star \mathcal{P}A$ also vanishes, as it is proportional to $\epsilon^{\alpha\beta}[\mathcal{P}A_{\alpha},\mathcal{P}A_{\beta}]=0$. Hence, the Lax curvature is simply proportional to the equations of motion
\begin{align}
\mathcal{F} L &=z \mathcal{E}.
\end{align}
\subsection{Integrable deformations}
In this subsection we explore the potential for an integrable deformation. Starting with the Lax connection \eqref{eq:undeformedLax}, we consider a slight generalization
\begin{align}
  \label{eq:genLax}
L(z)=J+ z \star \mathcal{P} I,
\end{align}
where $J$ and $I$ are algebra-valued one-forms. Due to the same Poincar\'e algebra property as before, the flatness condition is
\begin{align}
\mathcal{F}L=\mathcal{F}J+z \left( d \star \mathcal{P} I -  J \wedge  \star \mathcal{P}I -\star \mathcal{P}I \wedge J \right).
\end{align}
This implies that the Lax connection is suitable for a model with the equations of motion
\begin{align}
\mathcal{E} =d \star \mathcal{P} I -  J \wedge  \star \mathcal{P}I -\star \mathcal{P}I \wedge J,
\end{align}
provided $\mathcal{F}J =0$, at least on-shell. The Yang-Baxter model is such an example with $J=I$ and $I$ is defined by a deformation operator acting on $A$. In this case, the action yields precisely such equations of motion, however the curvature condition provides additional constraints in the form of the classical Yang-Baxter equation \cite{Klimcik:2008eq,Delduc:2013fga}.

In this paper, we focus on the case where $J=A$, leaving $I$ arbitrary for now. This means that the constraint $\mathcal{F} J =0$ is identically fulfilled. Our target is to construct an action such that
\begin{align}
  \label{eq:desiredForm}
 \partial_{\alpha}\mathcal{P} I^{\alpha}-[A_{\alpha},\mathcal{P} I^{\alpha}]=0
\end{align}
is satisfied only on-shell.
\section{Linear flat space deformations}
We introduce the deformed action based on the same Maurer-Cartan form $A=-g^{-1}dg$,
\begin{align}
  S[g]&=\frac12 \int_{\Sigma} \ip{A}{\star \mathcal{P} D A},\\
  D&=1+\frac{\kappa_{-}}{2} \left( K_{-}-K^{T}_{-} \right)\mathcal{P}\star + \frac{\kappa_{+}}{2} \left( K_{+}+K^{T}_{+} \right)\mathcal{P},
\end{align}
where the operators are inherently $g$-dependent to maintain coset gauge invariance and are defined by
\begin{align}
  K_{\pm}&=k_{\pm}^{ij}g^{-1}T_{i}g\ip{g^{-1}T_{j}g}{\bullet},\\
  K_{\pm}^{T}&=k_{\pm}^{ji}g^{-1}T_{i}g\ip{g^{-1}T_{j}g}{\bullet},
\end{align}
and $T_{i}$ are algebra elements.\footnote{This deformation operator is reminiscent of the Yang-Baxter deformation operator, in that case, the linear operator, $R_{g},$ is exclusively antisymmetric and is applied repeatedly as a geometric series, $D_{\mathrm{YB}}=\frac{1}{1+\eta R_{g} \mathcal{P} \star}$.}
It turns out, that it is more convenient to define the action in this manner rather than requiring $k_{-}^{ij}$ and $k_{+}^{ij}$ to be antisymmetric and symmetric, respectively.
The resulting equations of motion are
\begin{align}
\mathcal{E}=D_{\alpha}\mathcal{P}A^{\alpha}&+\frac{\kappa_{-}}{2} D_{\alpha}\mathcal{P}\left( K_{-}-K_{-}^{T} \right)\mathcal{P} \left( \star A\right)^{\alpha}+\frac{\kappa_{-}}{2} [\mathcal{P} A_{\alpha},\left( K_{-}-K_{-}^{T} \right) \mathcal{P} \left( \star A\right)^{\alpha}]\nonumber\\&+ \frac{\kappa_{+}}{2} D_{\alpha}\mathcal{P}\left(  K_{+}+K_{+}^{T}\right)\mathcal{P}A^{\alpha}+ \frac{\kappa_{+}}{2}[\mathcal{P} A_{\alpha},\left(  K_{+}+K_{+}^{T}\right)\mathcal{P} A^{\alpha}],
\end{align}
where $\left( \star A\right)^{\alpha}=\epsilon^{\alpha\beta}A_{\beta}$ and $D_{\alpha}=\partial_{\alpha}-[A_{\alpha},\bullet]$. This is not manifestly in the desired form \eqref{eq:desiredForm}. To proceed, we need the four commutator terms $(K_{\pm},K_{\pm}^{T})$ to either vanish or be rewritten as a total derivative, at least on-shell. For the commutator terms to vanish we require
\begin{align}
[K_{\pm}^{(T)}X,Y]\pm [K_{\pm}^{(T)}Y,X]=0,\quad  \forall X,Y\in\mathfrak{g}^{(1)}.
\end{align}
where $K^{(T)}_{\pm}$ can be either $K_{\pm}$ or $K^{T}_{\pm}$.\footnote{In a later section, we will see that if the commutator terms vanish off-shell, the translation symmetries remain undeformed. Consequently, the resulting backgrounds are limited to flat space with a constant $B$-field.} Alternatively, we can rewrite the commutators as total derivatives by employing chain rule identities\footnote{The following identities are derived from the chain rule $ \ip{T}{D_{\alpha} \mathcal{P}K \mathcal{P}A^{\alpha}_{\pm}}=  \ip{T}{\mathcal{P} K D_{\alpha}\mathcal{P}A^{\alpha}_{\pm}}-\ip{T}{[K \mathcal{P}A^{\alpha}_{\pm},\mathcal{P}A_{\alpha}]}+\ip{[\mathcal{P}A_{\alpha},K^{T}\mathcal{P}T]}{\mathcal{P}A^{\alpha}_{\pm}}, \forall T \in \mathfrak{g}$, where $K$ can be either $K_{\pm}$ and $A_{\pm}=\frac{1}{2} \left( 1\pm \star \right)A$.}
\begin{align}
\label{eq:8}
  D_{\alpha} \mathcal{P}K^{(T)}_{+} \mathcal{P}A^{\alpha}&= [\mathcal{P}A_{\alpha},K^{(T)}_{+} \mathcal{P}A^{\alpha}],
\end{align}
provided $\mathcal{P} K^{(T)}_{+} \mathcal{E} _{0}=0$, where
\begin{align}
\mathcal{E} _{0}=D_{\alpha}\mathcal{P}A^{\alpha},
\end{align}
are the undeformed equations of motion. For the $K_{-}$ terms, we have
\begin{align}
\label{eq:34}
D_{\alpha} \mathcal{P}K_{-} \mathcal{P} \left( \star A\right)^{\alpha}=[\mathcal{P}A_{\alpha},K_{-} \mathcal{P} \left( \star A\right)^{\alpha}],
\end{align}
provided $[X,K_{-}^{T}Y]=0,\forall X,Y\in \mathfrak{g}^{(1)}$. Since this condition implies that the $K_{-}^{T}$ commutator term vanishes, we are forced to have at least one of the $K_{-}$ terms vanish. In this paper, we consider the case where the $K_{\pm}^{T}$ commutator terms vanish and the $K_{\pm}$ terms are rewritten as total derivatives. For this, we can propose the following Lax connection:
\begin{align}
  L(z)&=A+z \star \mathcal{P} I,\\
  I&=A+ \frac{\kappa_{-}}{2} \left( 2 K_{-}-K_{-}^{T}  \right)\star\mathcal{P}  A+\frac{\kappa_{+}}{2} \left( 2 K_{+}+ K_{+}^{T} \right) \mathcal{P} A,
\end{align}
given the following on-shell conditions
\begin{align}\label{eq:conditions}
 & \mathcal{P} K_{+} \mathcal{E} _{0}=0, &
 & [K_{+}^{T}X,Y]+ [K_{+}^{T}Y,X]=0,&
&[X,K_{-}^{T}Y]=0,
\end{align}
for all $X,Y\in \mathfrak{g}^{(1)}$. In Appendix \ref{app:constraint} we analyze the on-shell constraint $\mathcal{P} K_{+} \mathcal{E} _{0}=0$ and express it as an algebraic operator constraint. For the purposes of this paper, we find solutions to this constraint by recognizing it is on-shell equivalent to $\mathcal{P} K_{+} \left( \mathcal{E}-\mathcal{E} _{0} \right)=0$.

\subsection{Coordinate form}
When introducing a coset representative and expressing the action in coordinate form,
\begin{align}
    S[x]=\frac12 \int_{\Sigma}  dx^{\mu} \wedge \left( \star G_{\mu\nu} -B_{\mu\nu} \right)dx^{\nu},
\end{align}
the operator $K_{+}$ will only deform the metric, while $K_{-}$ will only contribute to the $B$-field. The deformed geometry is given by
\begin{align}
  G_{\mu\nu}&= g_{\mu\nu}+\kappa_{+}k_{+}^{(ij)} \chi_{i \mu}\chi_{j\nu},\\
  B_{\mu\nu}&= - \kappa_{-} k_{-}^{[ij]}  \chi_{i \mu}\chi_{j\nu},
\end{align}
where $g_{\mu\nu}=\ip{A_{\mu}}{\mathcal{P}A_{\nu}}$ is the undeformed metric and $\chi_{i \mu}=\ip{g^{-1}T_{i}g}{\mathcal{P}A_{\mu}}$ are the components of the Killing vector corresponding to generator $T_{i}$, here $A_{\mu}=-g^{-1}\partial_{\mu}g$. The (anti-)symmetrization brackets are defined as $k^{[ij]}=\frac12 \left( k^{ij}-k^{ji} \right)$ and $k^{(ij)}=\frac12 \left( k^{ij}+k^{ji} \right)$.

\section{Examples}
In this section, we will explore examples of deformed backgrounds that can be achieved using this construction. For all the following cases, we will assume Cartesian coordinates by selecting a coset representative of the form
\begin{align}
g=e^{x^{\mu}p_{\mu}}.
\end{align}
\subsection{Constant $H$-flux flat space}
A straightforward way to satisfy the conditions \eqref{eq:conditions} is to have no metric deformation $K_{+}=0$, and let
\begin{align}
K_{-}= C^{\mu\nu\rho}g^{-1}m_{\mu\nu} g\ip{g^{-1}p_{\rho} g}{\bullet},
\end{align}
for arbitrary $C^{\mu\nu\rho}$. This allows us to achieve any desired constant $H=dB$
\begin{align}
H=C_{\mu\nu\rho}dx^{\mu}\wedge dx^{\nu}\wedge dx^{\rho},
\end{align}
where $C_{\mu\nu\rho}=g_{\mu\mu'}g_{\nu\nu'}g_{\rho\rho'}C^{\mu\nu\rho}$ . It turns out that constant $H$-flux is the best we can achieve when $K_{+}=0$, as translations remain off-shell symmetries up to a total derivative.
\subsection{Metric deforming solutions}\label{sec:metric}
We can also find metric deformations. To satisfy the constraint $[K_{+}^{T}X,Y]+ [K_{+}^{T}Y,X]=0$, we will consider
\begin{align}\label{eq:rank1}
K_{+}=g^{-1}M g\ip{g^{-1}P g}{\bullet},
\end{align}
with $M\in \mathfrak{g}^{(0)},P\in \mathfrak{g}^{(1)}$. We now need to satisfy the additional on-shell constraint that is $ \mathcal{P}K_{+}\mathcal{E}_{0} =0$. On-shell this constraint is equivalent to  $ \mathcal{P}K_{+}\left( \mathcal{E}-\mathcal{E}_{0} \right) =0$, which we will solve this instead. Assuming the Cartesian coset representative, to simplify our discussion, the constraint boils down to $\ip{P }{ \mathcal{E}-\mathcal{E}_{0}}=0$. This implies that a particular component of $\mathcal{E}=\mathcal{E}^{\mu}p_{\mu}$ remains undeformed. This can happen when using light cone directions, $p_{\pm}=\frac{p_{0}\pm p_{1}}{\sqrt{2}}$. For example, $\ip{p_{-}}{ \mathcal{E}-\mathcal{E}_{0}}=0$ requires only that the $\mathcal{E}^{+}$ component to remain undeformed, whilst a $K_{+}$ with $P=p_{-}$ is only deforming the $\mathcal{E}^{-}$ component and the directions affected by $M$. In conclusion, the conditions are satisfied provided that $M$ does not involve $m_{+\mu}$. This leaves a two-parameter deformation in four dimensions:
\begin{align}
K_{+}=g^{-1}\left(\kappa_{1}m_{-2} +\kappa_{2}  m_{23}\right) g\ip{g^{-1}p_{-}g}{\bullet}.
\end{align}
With this deformation operator, the background becomes
\begin{align}
G_{\mu\nu}dx^{\mu}dx^{\nu}&=\kappa_{1} \left(x^{2} dx^{+}-x^{+}dx^{2}  \right)dx^{+}+\kappa_{2} \left(x^{2} dx^{3}-x^{3}dx^{2}  \right)dx^{+}\nonumber\\&\quad -2 dx^{+}dx^{-}+dx^{2}dx^{2}+dx^{3}dx^{3}.
\end{align}
For $\kappa_{2}=0$, this background is simply flat space, as the deformation can then be removed by the coordinate transformation
\begin{align}\label{eq:transformation}
x^{+}=\tilde{x}^{+},&& x^{-}=\tilde{x}^{-}+\frac{\kappa_{1}}{2} \tilde{x}^{+}\tilde{x}^{2}+ \frac{\kappa_{1}^{2}}{12}\left( \tilde{x}^{+} \right)^{3},&& x^{2}=\tilde{x}^{2}+\frac{\kappa_{1}}{2} \left( \tilde{x}^{+} \right)^{2},&& x^{3}=\tilde{x}^{3}.
\end{align}
The fact that the background remains flat means that the deformation does not break any symmetries, as it remains maximally symmetric. We will explore the symmetries of these models in the next section. On the other hand, for $\kappa_{2}\neq 0$ we can eliminate the $\kappa_{1}$ term by the transformation
\begin{align}\label{eq:transformation2}
x^{+}=\tilde{x}^{+},&& x^{-}=\tilde{x}^{-}-\frac{\kappa_{1}}{\kappa_{2}} \tilde{x}^{3}+\frac{\kappa_{1}^{2}}{2\kappa_{2}^{2}} \tilde{x}^{+},&& x^{2}=\tilde{x}^{2},&& x^{3}=\tilde{x}^{3}-\frac{\kappa_{1}}{\kappa_{2}}\tilde{x}^{+}.
\end{align}
In conclusion, only the $\kappa_{2}$ deformation is physical.

We can also include a $K_{-}$ operator that does not deform the $p_{+}$ direction. In four dimensions, there is only one inequivalent option up to total derivatives. In summary, for the operators
\begin{align}
  K_{+}=g^{-1}m_{23} g\ip{g^{-1}p_{-} g}{\bullet},\\
  K_{-}=g^{-1}m_{23} g\ip{g^{-1}p_{-} g}{\bullet},
\end{align}
in Cartesian light cone coordinates $(x^{+},x^{-},x^{2},x^{3})$, the background comes out to be
\begin{align}
  \left( G+B \right)_{\mu\nu}=\mqty(
  0 & -1 & -\frac{1}{2}x^{3} \left( \kappa_{+}+\kappa_{-} \right) & \frac{1}{2}x^{2} \left(\kappa_{+}+\kappa_{-} \right) \\
  -1 & 0 & 0 & 0\\
  -\frac{1}{2}x^{3} \left( \kappa_{+}-\kappa_{-} \right) &0 &1& 0\\
  \frac{1}{2}x^{2} \left(\kappa_{+}-\kappa_{-} \right) &0 &0& 1).
\end{align}
For $\kappa_{+}=\kappa_{-}$, this is precisely the Nappi-Witten model \cite{Nappi:1993ie}. Curiously, the Nappi-Witten model can also be obtained as a Yang-Baxter deformation of flat space with the $r$-matrix $r= p_{-}\wedge m_{23}$\footnote{The Yang-Baxter prescription yields the background in a different coordinate system and provides a distinct Lax connection with a different set of manifest symmetries. The coordinate system presented here is related to the coordinates of \cite{Idiab:2024bwr} by the following transformation: $x^{+}=\tilde{x}^{+},x^{-}=\tilde{x}^{-},x^{2}= \tilde{x}^{2} \cos(\frac{\kappa_{+} }{2}\tilde{x}^{+})+ \tilde{x}^{3} \sin(\frac{\kappa_{+} }{2}\tilde{x}^{+}),x^{3}= \tilde{x}^{3} \cos(\frac{\kappa_{+} }{2}\tilde{x}^{+})- \tilde{x}^{2} \sin(\frac{\kappa_{+} }{2}\tilde{x}^{+})$  .} \cite{Idiab:2024bwr}, in that case, the $B$-field parameter is also fixed $\kappa_{+}=\pm \kappa_{-}$. However, a Yang-Baxter deformation of the Nappi-Witten model itself can rescale the $B$-field parameter and thus yield the same background as found here \cite{Kyono:2015iqs}. For unequal parameters the background may still solve the type IIB SUGRA equations of motion with a non-constant dilaton $\phi=\frac{1}{8}\left(\kappa_{-}^{2}-\kappa_{+}^{2}  \right)\left( x^{+} \right)^{2}$. The string coupling remains bounded for $\kappa_{-}^{2}\geq \kappa_{+}^{2}$.

\subsection{Extended Nappi-Witten model}
For our next example, the six dimensional generalization of the Nappi-Witten background, as described in \cite{Idiab:2024bwr}, can be obtained using the following operators
\begin{align}
  K_{+}&=g^{-1}\left( \kappa_{1} m_{23} +\kappa_{2} m_{45} \right) g\ip{g^{-1}p_{-} g}{\bullet},\\
  K_{-}&=g^{-1}\left( \eta_{1} m_{23} +\eta_{2} m_{45} \right) g\ip{g^{-1}p_{-} g}{\bullet}.
\end{align}
In light cone coordinates, the background is given by
\begin{align}
  G_{\mu\nu}dx^{\mu}dx^{\nu}&=\kappa_{1}\left(x^{2}dx^{3}- x^{3}dx^{2} \right)dx^{+}+\kappa_{2}\left(x^{4}dx^{5}- x^{5}dx^{4} \right)dx^{+} \nonumber\\&\quad-2dx^{+}dx^{-}+dx_{i} dx^{i},\\
  B&= \frac{\eta_{1}}{2} \left(x^{3} dx^{2}-x^{2} dx^{3}  \right)\wedge dx^{+}+\frac{\eta_{2}}{2} \left(x^{5} dx^{4}-x^{4} dx^{5}  \right)\wedge dx^{+},
\end{align}
having ignored the overall deformation parameters by setting $\kappa_{\pm}=1$. This setup allows for the freedom to scale the $B$-field $\kappa_{i}\neq \eta_{i}$ compared to the Yang-Baxter model constructed in \cite{Idiab:2024bwr}. In fact, in the six dimensional case, there is even more freedom because there are additional inequivalent $K_{-}$ operators that do not include the $p_{+}$ direction
\begin{align}
K_{-}= C^{ijk }g^{-1}m_{ij} g\ip{g^{-1}p_{k} g}{\bullet},
\end{align}
with the indices running over $i,j,k\in\{-,2,3,4,5\}$. This yields any constant $H$-flux that does not involve $dx^{-}$,
\begin{align}
 H&=C_{ijk} dx^{i}\wedge dx^{j}\wedge dx^{k},
\end{align}
here the indices are lowered and therefore run over $i,j,k\in\{+,2,3,4,5\}$. In other words, we can obtain Nappi-Witten-like models with nearly arbitrary constant $H$-flux.


\subsection{Non-constant $H$-flux}\label{sec:nonlinear}
In the analysis of Section \ref{sec:metric}, we observed a metric deformation resulting in flat space in a different coordinate system. The coordinate transformation \eqref{eq:transformation} is non-linear and can be combined with a $B$-field deformation to produce a non-constant $H$-flux upon transforming back to Cartesian coordinates.\footnote{This will not work for the Nappi-Witten models $(\kappa_{2}\neq 0)$ as the corresponding transformation \eqref{eq:transformation2} is linear.} For this, we require at least five dimensions. Consider the following operators:
\begin{align}
\label{eq:4}
K_{+}&=g^{-1}m_{-2} g\ip{g^{-1}p_{-} g}{\bullet},\\
  K_{-}&=g^{-1}m_{23} g\ip{g^{-1}p_{4} g}{\bullet}.
\end{align}
After applying the coordinate transformation \eqref{eq:transformation}, this setup yields flat space with a non-constant $H$-flux of the form
\begin{align}
H= \kappa_{-} dx^{2}\wedge dx^{3}\wedge dx^{4} +  \kappa_{-}\kappa_{+} x^{+} dx^{+}\wedge dx^{3}\wedge dx^{4}.
\end{align}

\section{Conserved charges and Lagrangian symmetries}
In the undeformed model, symmetries correspond to left multiplications of constant group elements, $g\rightarrow k g $. For infinitesimal algebra elements, this translates to $g\rightarrow \left( 1+\epsilon \right)g, \epsilon\in \mathfrak{g}$. These symmetries remain in the deformed model if they commute with the deformation operator, to be explicit, they should solve
\begin{align}
 \ip{\mathcal{P}A}{\left[  \kappa_{-}  \left(K_{-}-K^{T}_{-}\right)\mathcal{P} A+\kappa_{+} \left(K_{+}+K^{T}_{+}\right)\star\mathcal{P}  A, g^{-1}\epsilon g\right] }=0.
\end{align}
Deformed or enhanced symmetries may also arise. These correspond to Lagrangian symmetries by non-constant infinitesimal variations $g\rightarrow \left( 1+\epsilon \right)g, d\epsilon \neq 0$. If the non-constant $\epsilon$ is deformation parameter dependent and reverts to a Poincar\'e symmetry in the undeformed limit, we call it a deformation of that Poincar\'e symmetry. Generally, these (deformed) symmetries can be difficult to identify, but give rise to useful conserved currents. From the Lax connection, we can also construct conserved currents. Consider the gauge transformation\footnote{Having a Lax connection means we can find another one by transforming it as $L\rightarrow L'=h L h^{-1}+dh h^{-1}$ for any $h$. This transformation has the property that $\mathcal{F}L'=h \left( \mathcal{F} L \right)h^{-1}$ and therefore $\mathcal{F}L = 0 \iff \mathcal{F} L' =0$. }
\begin{align}
L'&=g L g^{-1}+dg g^{-1},
\end{align}
where $g$ is the coset representative. Recalling the original Lax connection $L=A + z\star \mathcal{P} I$, the transformation cancels the $A$ term and therefore $L'\in \mathfrak{g}^{(1)}$.
Having a Lax connection entirely in $\mathfrak{g}^{(1)}$ implies $L' \wedge L' =0$ and the curvature condition simplifies to $d L' =0$, yielding an on-shell conserved current
\begin{align}
d \left[  g \left( \star \mathcal{P}I  \right) g^{-1} \right]= 0.
\end{align}
The current being in grade $1$ and non-degenerate in the undeformed limit, implies we have the same number of conserved currents as the target space dimension. Additionally, the current is local and may correspond to Lagrangian symmetries. Comparing to the undeformed Noether current
\begin{align}
d\ip{g^{-1}T g}{\star \mathcal{P} A} =0,
\end{align}
with $T\in \mathfrak{g}$, in the undeformed case, the conserved Lax current corresponds to the Noether currents for translations
\begin{align}
p^{\mu}\ip{g^{-1}p_{\mu} g}{\star \mathcal{P} A}=g \left( \star \mathcal{P}A  \right) g^{-1}.
\end{align}
Thus, we conclude that for our deformations, the translation symmetries are never fully broken but may be deformed.\footnote{This line of reasoning does not apply to Yang-Baxter deformations, as the gauge transformation required to convert the Lax connection exclusively into $\mathfrak{g}^{(1)}$ is not necessarily local.} For all the examples considered in the previous section, we find that in Cartesian coordinates, the deformed translation Killing vectors are given by $\chi^{\mu}p_{\mu}=\left( 1+\frac{\kappa_{+}}{2}K_{+}^{T} \right)c^{\mu}p_{\mu}$, which means $\mathcal{L}_{\chi} \left( g+B \right) =dC$.


\section{Conclusions and outlook}

In this paper, we have introduced a deformation of the flat space symmetric space sigma model by incorporating a gauge-compatible linear operator into the action. We identify certain conditions on the operator such that we can find a Lax connection whose flatness condition is equivalent to the equations of motion. Then, we explored simple solutions to the constraint and examined their resulting background.

It turns out that any constant $H$-flux in flat space, expressed in Cartesian coordinates, can be obtained by this type of deformation. When searching for metric deforming solutions we only manage to achieve the Nappi-Witten and extended Nappi-Witten metric with the option of adding any constant $H$-flux that does not involve the $dx^{-}$ direction. In four dimensions this corresponds to rescaling the $B$-field, a transformation that can also be achieved through a Yang-Baxter deformation of the Nappi-Witten model itself \cite{Kyono:2015iqs}. It is an intersting question whether the additional parameters allowed for the six dimensional case can similarly be obtained through Yang-Baxter deformations.

In examining the symmetries of the deformed backgrounds, we find that this type of deformation consistently retains a set of symmetries that revert to the translation symmetries of flat space in the undeformed limit. For the deformed backgrounds to differ from a maximally symmetric space, they must hence break Lorentz symmetries. Some deformation operators fail to break, but only deform these symmetries, resulting in backgrounds that remain flat space. Determining the precise conditions under which Lorentz symmetries are either fully broken or preserved is an open question. It would be beneficial to derive explicit formulas for the Killing vectors or conserved charges associated with the deformed Lorentz symmetries. On the other hand, conserved currents associated to the deformed translation algebra are easily derived from the Lax connection.

A number of open questions remain. For instance, it would be interesting to understand the complete space of deformations. At the moment, it is unclear to what extent further $H$-flux deformations are possible and whether metric deformations beyond extended Nappi-Witten models are possible.
To further explore this issue, closer examination of the on-shell constraint $K_{+}\mathcal{E}_{0}=0$ is necessary. It would be interesting to relax the assumption \eqref{eq:rank1} and look for higher rank solutions.

While we have demonstrated that the system admits a Lax connection, establishing its complete integrability remains an open challenge. However, many of the deformed backgrounds that we demonstrated correspond to known integrable systems from other formulations \cite{Nappi:1993ie,Kyono:2015iqs,Idiab:2024bwr}. It would be interesting to study the worldsheet $S$-matrices and Hamiltonians for these systems. It is plausible that all these deformations can be gauge-fixed such that the worldsheet Hamiltionian is quadratic. Furthermore, the potential of these deformed backgrounds to solve the SUGRA equations of motion is an unanswered question. It is likely that Weyl invariance would place further constraints on the deformations, similar to the case for Yang-Baxter deformation \cite{Borsato:2016ose,Hronek:2020skb}.

More broadly, it would be valuable to experiment with various actions that yield equations of motion compatible with the Lax connection \eqref{eq:genLax}, whether $J=A$ or $J\neq A$. Finally, extending this framework to $\mathrm{AdS}_{5}$ could be particularly relevant in the context of AdS/CFT, or more generally, for semi-simple symmetric spaces.

\section*{Acknowledgements}
The author expresses gratitude to Stijn J. van Tongeren and Ben Hoare for their insightful discussions and valuable suggestions. Special thanks also to Stijn J. van Tongeren for his helpful comments on the draft. This work is supported by the German Research Foundation (DFG) via the Emmy Noether program “Exact Results in Extended Holography”.

\appendix
\section{On-shell constraint in algebraic operator form}\label{app:constraint}
In this section we analyze the on-shell constraint $K_{+} \mathcal{E}_{0}=0$. We start by writing the equations of motion in the following form
\begin{align}
\mathcal{E}- \mathcal{E}_{0}&=\frac{\kappa_{-}}{2} D_{\alpha}\mathcal{P}\left( K_{-}-K_{-}^{T} \right)\mathcal{P} \left( \star A\right)^{\alpha}+\frac{\kappa_{-}}{2} [\mathcal{P} A_{\alpha},\left( K_{-}-K_{-}^{T} \right) \mathcal{P} \left( \star A\right)^{\alpha}]\nonumber\\&+ \frac{\kappa_{+}}{2} D_{\alpha}\mathcal{P}\left(  K_{+}+K_{+}^{T}\right)\mathcal{P}A^{\alpha}+ \frac{\kappa_{+}}{2}[\mathcal{P} A_{\alpha},\left(  K_{+}+K_{+}^{T}\right)\mathcal{P} A^{\alpha}],
\end{align}
next, we carry out the derivatives using the chain rule
\begin{align}
\mathcal{E}-\mathcal{E}_{0} &=\frac{\kappa_{-}}{2} (k^{ij}_{-}-k^{ji}_{-}) \mathcal{P}[g^{-1}T_{i}g] \ip{[\mathcal{P}A_{\alpha},g^{-1}T_{j}g]}{\mathcal{P}(\star A^{\alpha})}+\frac{\kappa_{+}}{2} \mathcal{P} \left( K_{+} +K_{+}^{T}\right) \mathcal{E}_{0}\nonumber\\&+\kappa_{-} [\mathcal{P} A_{\alpha},\left( K_{-}-K_{-}^{T} \right) \mathcal{P} \left( \star A\right)^{\alpha}]+ \kappa_{+}[\mathcal{P} A_{\alpha},\left(  K_{+}+K_{+}^{T}\right)\mathcal{P} A^{\alpha}].
\end{align}
On-shell we have $\mathcal{E} = 0$, which gives us the on-shell equation
\begin{align}
-&\left( 1+\frac{\kappa_{+}}{2} \mathcal{P} \left( K_{+} +K_{+}^{T}\right) \right)\mathcal{E}_{0} \approx\frac{\kappa_{-}}{2} (k^{ij}_{-}-k^{ji}_{-}) \mathcal{P}[g^{-1}T_{i}g] \ip{[\mathcal{P}A_{\alpha},g^{-1}T_{j}g]}{\mathcal{P}(\star A^{\alpha})}\nonumber\\&\quad  +\kappa_{-} [\mathcal{P} A_{\alpha},\left( K_{-}-K_{-}^{T} \right) \mathcal{P} \left( \star A\right)^{\alpha}]+ \kappa_{+}[\mathcal{P} A_{\alpha},\left(  K_{+}+K_{+}^{T}\right)\mathcal{P} A^{\alpha}],
\end{align}
where $\approx$ signifies on-shell equality. The operator acting on $\mathcal{E}_{0}$ can be inverted
\begin{align}
-\mathcal{E}_{0} &\approx\frac{\kappa_{-}}{2} (k^{ij}_{-}-k^{ji}_{-}) \left( 1+\frac{\kappa_{+}}{2} \mathcal{P} \left( K_{+} +K_{+}^{T}\right) \right)^{-1}\mathcal{P}[g^{-1}T_{i}g] \ip{[\mathcal{P}A_{\alpha},g^{-1}T_{j}g]}{\mathcal{P}(\star A^{\alpha})}\nonumber\\&+\kappa_{-} \left( 1+\frac{\kappa_{+}}{2} \mathcal{P} \left( K_{+} +K_{+}^{T}\right) \right)^{-1}[\mathcal{P} A_{\alpha},\left( K_{-}-K_{-}^{T} \right) \mathcal{P} \left( \star A\right)^{\alpha}]\nonumber\\&+ \kappa_{+}\left( 1+\frac{\kappa_{+}}{2} \mathcal{P} \left( K_{+} +K_{+}^{T}\right) \right)^{-1}[\mathcal{P} A_{\alpha},\left(  K_{+}+K_{+}^{T}\right)\mathcal{P} A^{\alpha}].
\end{align}
Now we have an on-shell expression of $\mathcal{E}_{0}$ and can find the components of $K_{+} \mathcal{E}_{0}$
\begin{align}
- \ip{T}{K_{+}\mathcal{E}_{0}} &\approx\frac{\kappa_{-}}{2}\ip{[\mathcal{P}A_{\alpha},\left( K^{T}_{-}-K_{-} \right)\mathcal{P} \frac{1}{1+\frac{\kappa_{+}}{2}  \left( K_{+} +K_{+}^{T}\right) \mathcal{P}}K^{T}_{+}T]}{\mathcal{P}(\star A^{\alpha})}\nonumber\\&+\kappa_{-} \ip{T}{K_{+} \frac{1}{1+\frac{\kappa_{+}}{2} \mathcal{P} \left( K_{+} +K_{+}^{T}\right)} [\mathcal{P} A_{\alpha},\left( K_{-}-K_{-}^{T} \right) \mathcal{P} \left( \star A\right)^{\alpha}]}\nonumber\\&+ \kappa_{+} \ip{T}{K_{+} \frac{1}{1+\frac{\kappa_{+}}{2} \mathcal{P} \left( K_{+} +K_{+}^{T}\right)}[\mathcal{P} A_{\alpha},\left(  K_{+}+K_{+}^{T}\right)\mathcal{P} A^{\alpha}]},
\end{align}
for arbitrary generator $T$. For our system we also require
\begin{align}
 & [K_{+}^{T}X,Y]+ [K_{+}^{T}Y,X]=0,&
&[X,K_{-}^{T}Y]=0,
\end{align}
for all $X,Y\in \mathfrak{g}^{(1)}$. We can use this to simplify and write the $\mathcal{P} K_{+} \mathcal{E} _{0}=0$ condition as the following algebraic conditions in operator form
\begin{align}
  \label{eq:38}
  &\mathcal{O}=\frac{1}{1+\frac{\kappa_{+}}{2}  \left( K_{+} +K_{+}^{T}\right)\mathcal{P}},\\\label{eq:kplus}
  &\mathcal{P} K_{+}  \mathcal{P} \mathcal{O} \left(   [X,K_{+}Y]+[Y,K_{+}X] \right) = 0,\\\label{eq:kplusminus}
  & \ip{ \mathcal{O}K_{+} ^{T}Z}{  [X,K_{-} Y]-[Y,K_{-} X] }=\ip{[X,K_{-}\mathcal{P} \mathcal{O}K^{T}_{+}Z]}{Y},
\end{align}
for all $X,Y,Z\in \mathfrak{g}^{(1)}$.


\bibliographystyle{elsarticle-num}
\bibliography{bibfile}

\end{document}